# Molecular Dynamics Study of the Catalyst Particle Size Dependence on Carbon Nanotube Growth

Feng Ding[*], Arne Rosén and Kim Bolton

Experimental Physics, School of Physics and Engineering Physics, Göteborg University and Chalmers University of Technology, SE-412 96, Göteborg, Sweden

## Abstract

The molecular dynamics method, based on an empirical potential energy surface, was used to study the effect of catalyst particle size on the growth mechanism and structure of single-walled carbon nanotubes (SWNTs). The temperature for nanotube nucleation (800-1100 K), which occurs on the surface of the cluster, is similar to that used in catalyst chemical vapor deposition experiments, and the growth mechanism, which is described within the vapor-liquid-solid model, is the same for all cluster sizes studied here (iron clusters containing between 10 and 200 atoms were simulated). Large catalyst particles, that contain at least 20 iron atoms, nucleate SWNTs and have a far better tubular structure than SWNTs nucleated from smaller clusters. In addition, the SWNTs that grow from the larger clusters have diameters that are similar to the cluster diameter, whereas the smaller clusters, which have diameters less than 0.5 nm, nucleate nanotubes that are approximately 0.6-0.7 nm in diameter. This is in agreement with the experimental observations that SWNT diameters are similar to the catalyst particle diameter, and that the narrowest free-standing SWNT is 0.6-0.7 nm.

*Corresponding author: Email-fengding@fy.chalmers.se
Tel- +46-31-7723294
On leave from Department of Physics, Qufu Normal University, Qufu, 273165, Shandong, P. R. China

# I. Introduction

Since their discovery in 1991,[1] carbon nanotubes (CNTs) have been the focus of much scientific interest because of their unique physical and chemical properties, as well as their potential technological applications such as hydrogen storage,[2] electronic devices[3] and chemical sensors.[4] Over the past decade there has been significant progress in CNT production methods, such as the mass production of well-aligned multi-walled carbon nanotubes (MWNTs),[5] the growth of high quality single-walled[6] carbon nanotubes (SWNTs), the production of SWNT bundles that are several decimeters long[7] and the growth of SWNTs that are millimeters in length.[8] In spite of this tremendous progress, the CNT growth mechanism is still poorly understood.

It is believed that the transition metal catalyst (*e.g.*, Fe, Co, Ni or their alloys) plays a key role in the growth of high quality, long CNTs. The catalyst particles, which are typically between 1 and 100 nm in diameter, can be on a substrate[9] or can 'float' in a gas[10]. The role of these particles in the growth mechanism is described in the widely accepted vapor-liquid-solid (VLS) model,[11] which was first proposed to explain the growth of carbon nanofibers in 1972.[12] According to this model, carbon that is evaporated from graphite or decomposed from carbon-rich gases (*e.g.*, CO, $C_6H_6$ and $C_2H_2$) dissolves into the liquid metal clusters, and when the metal-carbide clusters are supersaturated in carbon, carbon islands precipitate on the cluster surface and nucleate CNTs. Although the VLS model provides a simple explanation that captures the essence of CNT growth, it does not include atomistic details such as the mechanism of graphitic cap nucleation on the cluster surface, how the open end of the nanotube is maintained during CNT growth, how defects that may form in the CNT structure are healed, and what determines the diameter and chirality of the

CNT. This poor understanding of the detailed growth mechanism hinders further progress in CNT production, such as the selective growth of nanotubes that have a specific diameter or chirality.

Recent molecular dynamics (MD) studies have provided a deeper understanding of the atomistic details of catalyzed CNT growth. Simulations where the interatomic forces were obtained directly from DFT[13] showed that dissolved carbon precipitates from the metal-carbide cluster to form a heptagon and pentagon on the cluster surface. The addition of carbon atoms to the end of a growing SWNT was also simulated. Obtaining the forces directly from DFT calculations at each trajectory step ensures high accuracy, but also limits the system size (less than 200 atoms were simulated) and the number and length of the trajectories (tens of ps were simulated). The nucleation of the graphitic cap was thus not simulated in these studies. Simulations that are based on empirical potential energy surfaces (PESs)[14,15] cannot capture the time dependence of the electron density, and the catalytic role of the transition metal cluster has thus not been studied. However, the role of the metal particle as a solvent and a template for CNT growth can be studied using these PESs. In addition, an ensemble of long trajectories that contain many atoms can be propagated. For example, the pioneering simulations by Maruyama and coworkers[14] showed that carbon can be dissolved in Ni clusters before precipitating to form SWNTs. As discussed below, we have developed an empirical PES that correctly reproduces the trends in the iron-carbide (FeC) phase diagram (that may be important for the precipitation of carbon from saturated iron-carbide clusters) and the decreasing cluster melting point with decreasing cluster size (which may be important in determining the phase of the catalyst particle at CNT growth temperatures)[16]. Simulations based on this PES[15] show that

SWNTs nucleate on a $Fe_{50}C$ surface between 800 and 1400 K, which is similar to the temperature range needed to nucleate CNTs in catalyzed chemical vapor deposition (CCVD) experiments. The simulations also revealed atomistic details of the nucleation mechanism. In this contribution we use MD simulations based on the same PES to study SWNT growth from FeC clusters that contain between 10 and 200 Fe atoms.

## 2. Potential energy surface and simulation methods

The PES used in this work has been detailed elsewhere,[15] and is briefly discussed here for the sake of completeness. A key aspect of the PES is that it distinguishes between the dissolved carbon atoms (inside the metal cluster) and the precipitated carbon atoms (on the cluster surface). The interaction between the dissolved C atoms is described by a Lennard-Jones 12-6 potential, which serves primarily to keep the C atoms dispersed in the iron solvent. The interaction between the precipitated atoms is described by the Brenner potential,[17] which has been used previously to study CNT dynamics, including SWNT growth from metal particles.[14] The Fe-Fe interactions are described by many-body potentials which are based on the second moment approximation of the tight binding model,[19] and which are known to be suitable for studying the thermal properties of pure[18] and alloy[20] transition metal systems. The interaction between the Fe atoms and the dissolved C is described by the Johnson potential where the parameters have been fit to experimental iron-carbide data.[21] This potential has been used previously to study C dissolved in liquid iron.[22] The interaction between the Fe atoms and the precipitated C is also described by the Johnson potential, but the interaction strength depends on whether the C atom is in the central part of the SWNT or graphitic island (where it is bond-saturated with other precipitated C atoms) or whether it is at the end/edge of the nanotube or island. Distinguishing these types of precipitated atoms is important for the nucleation mechanism, and it based on DFT calculations[23, 24] that show that there is approximately an order of magnitude stronger interaction between bond-unsaturated carbon and Fe atoms than between bond-saturated carbon and Fe atoms.

The PES described above has been used to study the thermal properties of FeC cluster systems[15]. MD simulations based on this PES yielded the correct trends of the FeC phase diagram, where a eutectic was found at 20% carbon content by weight. In addition, the correct $N^{-1/3}$ dependence of the cluster melting point on the number of atoms, N, in the cluster was obtained. These results support the validity of using this PES to study the thermal dynamics of FeC systems, and yields correct trends (phase behavior and melting point dependence on cluster size) that are important in the VLS description of SWNT growth.

The trajectories were initialized by annealing the pure $Fe_N$ cluster, where N is the number of atoms, to its minimum energy structure before heating it to the desired temperature. The cluster was propagated at this temperature for 100 ps to ensure thermal equilibrium, after which carbon atoms are inserted to the central part of the cluster. It should be noted that addition of carbon atoms to the cluster surface, instead of to the cluster center as done here, does not significantly effect the growth mechanism or SWNT structure. The rate of carbon atom insertion was varied from 1 atom per 10 ps to 1 atom per 100 ps. This carbon insertion rate is more than four orders of magnitude larger than the experimental rate, but is required for the simulation of SWNT nucleation within a reasonable computational time (several trajectories were propagated for each cluster size to ensure the statistical convergence of the results). As discussed in a previous paper,[15] this large insertion rate (which can be viewed as a high carbon feedstock pressure) may result in many defects in the simulated SWNT structure, but the growth mechanism is expected to be the same as that found under experimental growth conditions. The Berendsen velocity scaling method[25] was used to maintain the cluster at the desired temperature.

## 3. Results and discussion

### 3.1. SWNT nucleation on Fe$_N$C clusters for N=50, 100 and 150

Fig.1 shows the nucleation of a SWNT on the surface of a Fe$_{150}$C cluster. The growth mechanism is similar to that found for the Fe$_{50}$C cluster, which has been discussed in a previous contribution[15]. It is also similar to the growth mechanism for all clusters containing more than about 20 Fe atoms, and is summarized with reference to Figs. 1 and 2 below.

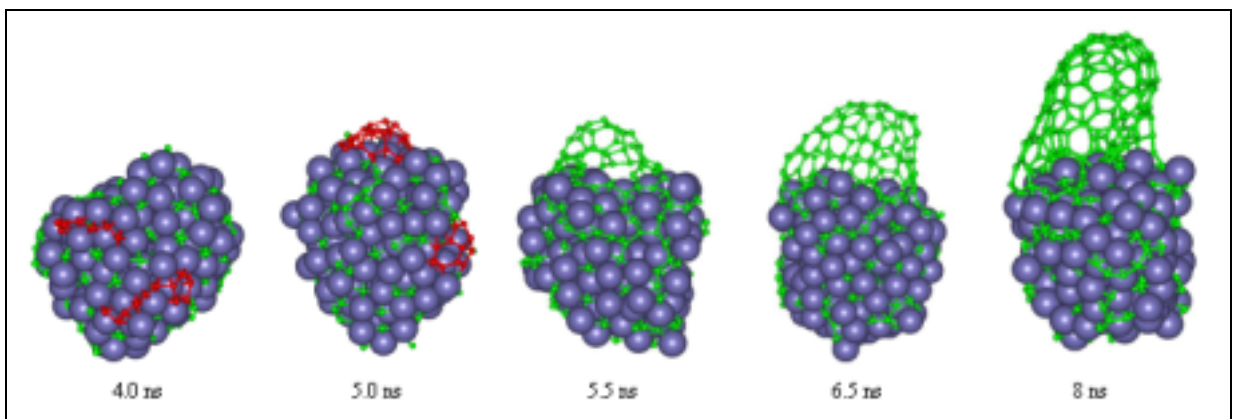

**Figure 1.** Nucleation of a carbon nanotube on the surface of a Fe$_{150}$ cluster. The temperature is 1100 K and a carbon atom is inserted into the cluster every 30 ps. The small dark and light gray spheres are carbon atoms and the large gray spheres are iron atoms.

The solid line in Fig. 2 shows the time dependence of the dissolved carbon concentration in the $Fe_{150}$ cluster, and is obtained from the same trajectory that yields the structures in Fig. 1. It is evident that there is a peak in the dissolved C concentration, which divides the nucleation into three distinct stages: 1) The unsaturated stage (t < 2.4 ns) where the cluster is unsaturated in carbon atoms (all carbon is dissolved), 2) the highly supersaturated stage (2.4 ns < t < 4.9 ns) when the dissolved carbon concentration is largest, and 3) the supersaturated stage (t > 4.9 ns) where the dissolved carbon concentration remains constant (and is slightly larger than the carbon content in $Fe_3C$).

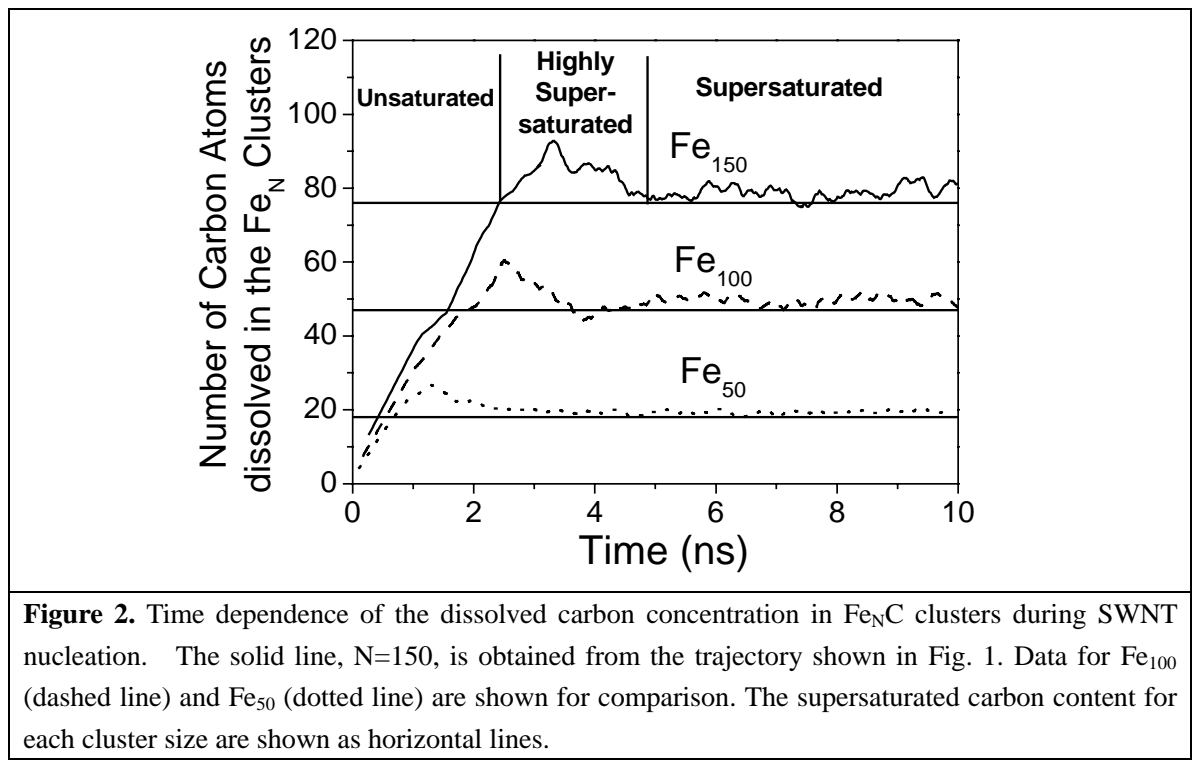

**Figure 2.** Time dependence of the dissolved carbon concentration in $Fe_NC$ clusters during SWNT nucleation. The solid line, N=150, is obtained from the trajectory shown in Fig. 1. Data for $Fe_{100}$ (dashed line) and $Fe_{50}$ (dotted line) are shown for comparison. The supersaturated carbon content for each cluster size are shown as horizontal lines.

During the first stage all of the carbon atoms that are inserted into the $Fe_{150}$ cluster are dissolved. The dissolved carbon concentration continues to increase even when the cluster is saturated in carbon, and a highly supersaturated cluster is obtained (the peak in Fig. 2). During the initial part of this supersaturated stage (stage 2), carbon atoms precipitate on the surface but dissolve back into the cluster. Only when the maximum dissolved carbon concentration is reached (≈3.5 ns) do the precipitated carbon atoms begin to nucleate carbon strings or polygons (the dark gray atoms in Fig. 1A) instead of dissolving back into the cluster. At this stage the dissolved carbon concentration begins to decrease. The carbon strings or polygons are more stable than isolated carbon atoms (which can dissolve back into the cluster) and they nucleate the formation of larger graphitic islands (shown in dark gray in Fig. 1B) Once the islands have a diameter of about 0.6 nm they can lift off the cluster surface (given that the temperature is sufficiently large so that the kinetic energy can overcome the attraction between the graphitic island and the cluster). Islands of this size contain several polygons (mainly heptagons and pentagons) so that there are sufficiently many bond-saturated C atoms that interact fairly weakly with the cluster atoms. Once the carbon island lifts off the cluster surface – to form a graphite cap – the dissolved carbon concentration remains constant at a supersaturated level (stage 3). During this stage the carbon atoms precipitate at the same rate at which they are inserted into the cluster, and they join at the open end of the graphite cap which increases in diameter and length. The diameter of the cap reaches a maximum when it is the same as that of the cluster, and at this stage all precipitating C atoms increase the length of the SWNT.

Although more than one graphitic island can form on the cluster surface, the larger clusters are more stable than the smaller ones. The smaller ones thus dissolve back into the cluster while the larger one forms the graphitic cap (see Figs. 1B and C).

The similarity of the solid, dashed and dotted curves in Fig. 2 shows that the SWNT nucleation mechanism does not depend on the size of the catalyst particle for N=50, 100 and 150. In all cases the cluster is highly supersaturated in carbon before carbon strings and polygons are nucleated, and the supersaturated dissolved carbon content during SWNT growth (stage 3) is constant. Since the cluster melting point depends on cluster size, the data in Fig. 2 were obtained at different temperatures for the various FeC clusters (about 900 K for N=150, 800 K for N=100 and 650 K for N=50). However, as discussed previously,[15] moderate changes in the temperature do not affect the SWNT growth mechanisms.

### 3.2. Effect of metal cluster size on SWNT structure and diameter

It is known experimentally that controlling the catalyst cluster size can assist in controlling the CNT structure and diameter. MWNTs[26] and SWNTs[27] that grow from small catalyst clusters (≈1 nm in diameter) have similar diameters as the catalyst particle. Also, experimental data suggests that the thinnest free-standing SWNT that grows catalytically is 0.6-0.7 nm[28, 29] (zeolite templates are required to grow 0.4 nm SWNTs[30]). The 0.6-0.7 nm lower limit for free standing nanotubes may result from the high curvature energy, *i.e.*, the curvature energy for 0.6 nm SWNTs is approximately 0.2 eV/atom, which is comparable to the thermal kinetic energy of about 0.12-0.19 eV/atom under typical CCVD conditions (900 – 1500 K), and approximately 0.6 eV/atom for 0.4 nm SWNTs.[31] If one assumes that the diameter of the free standing SWNT is the same as the catalyst particle diameter, then the 0.6-0.7 nm nanotubes grow from clusters that contain about 20 atoms (the diameter of Co/Ni/Fe atoms is about 0.25-0.27 nm). In agreement with this, recent experimental results show that SWNTs can grow from Co clusters that contain just 20-30 atoms.[32]

Fig. 3 shows typical simulated SWNT structures that are obtained from Fe particles that contain between 10 and 100 atoms. In order to isolate the effect of the cluster size, it was attempted to have the same growth conditions for all clusters. However, small variations (that do not affect the conclusions) were required due, for example, to the dependence of the cluster melting point on its size. Hence, a carbon atom was inserted into the $Fe_NC$ cluster every 80 (N=10, 15, 20 and 30) or 40 ps (N=50 and 100), and the simulation temperature was 800 (N=10, 15 and 20), 900, (N=30 and 50) or 1000 K (N=100). The structure grown from the $Fe_{10}$ cluster (Panel A) has the worst structure with many dangling bonds (*i.e.*, the carbon atoms are bond-unsaturated), and it is difficult to identify which carbon atoms are part of the tubular surface and which are on the inside (*i.e.*, for larger clusters these atoms are distinguished as dark or light gray spheres). The structure grown from the $Fe_{15}$ cluster (Panel B) is better than that grown from $Fe_{10}$, although there are still some dangling bonds and the nanotube diameter changes during the growth process (seen as large 'dents' in the tubular structure). The tubular structure of the SWNTs improve dramatically for clusters that contain at least 20 atoms. It is evident from Figs. 3C-F that, compared to growth from small clusters, SWNTs grown from clusters containing 20-100 Fe atoms have far better tubular structures, the diameters of these structures are constant during the growth, and very few dangling bonds are present.

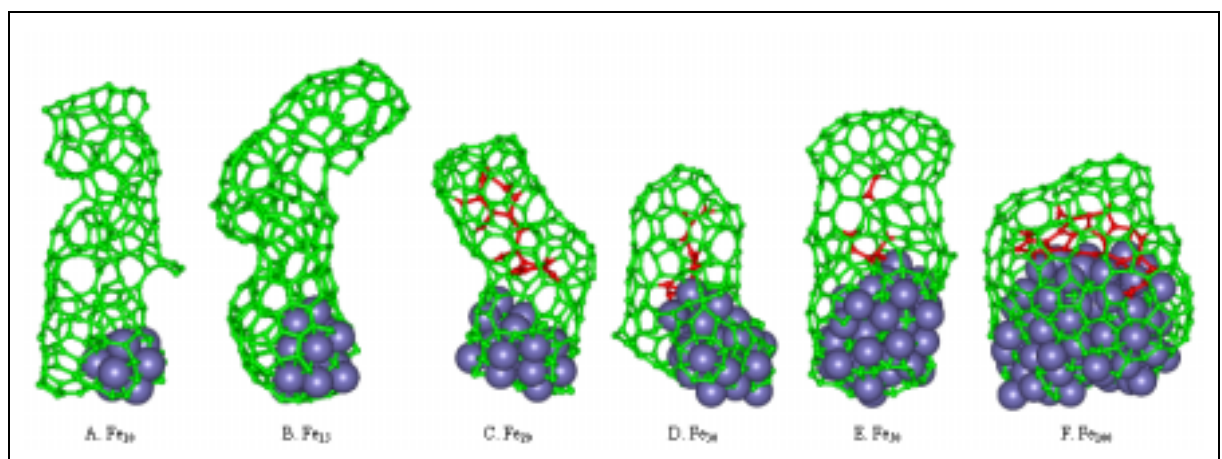

**Figure 3.** Dependence of the SWNT structure and diameter on cluster size. Panels A-F show typical structures that were obtained from clusters containing 10, 15, 20, 30, 50 and 100 Fe atoms, respectively.

Carbon atoms that form part of the tubular surface are shown in light gray, carbon atoms inside the tube are shown in dark gray and iron atoms are large gray spheres.

It is also evident from Fig. 3 that, in addition to the larger clusters (N≥20) producing SWNTs with the best tubular structures, these clusters also produce SWNTs that have diameters similar to the cluster diameter. Hence, in agreement with the experimental results mentioned above, the thinnest, well-structured SWNT grows from the $Fe_{20}$ cluster, which has a diameter of ≈0.6-0.7 nm (since Ni and Co have similar diameters to Fe, the smallest Ni and Co clusters that produce well-structured 0.6-0.7 nm SWNTs will also contain about 20 atoms). The simulated data shown in Figs. 3A and B also indicate that clusters with diameters smaller than 0.6-0.7 nm do not produce thinner nanotubes, but that the SWNTs are still 0.6-0.7 nm in diameter. However, as discussed above, these SWNTs have many defects and do not have a diameter that is constant over the length of the nanotube. The poor structure of these nanotubes is due to the fact that the $Fe_{10}$ and $Fe_{15}$ clusters are not sufficiently large to entirely cover the open end of the growing nanotube (see Figs. 3A and B). As a result, addition of carbon atoms at the open end is not uniform around the nanotube circumference, but growth only occurs at the part of the nanotube wall that is in contact with the metal particle. At a later stage in the growth process the other regions of the nanotube wall come into contact with the metal particle which allows for the addition of more carbon atoms in these regions. This non-uniformity in the addition of carbon atoms increases with decreasing particle size, and hence more defects appear in the SWNT cluster.

The relationship between the SWNT and cluster diameters that is discussed above is explicitly shown in Fig. 4. Except for $Fe_{10}$ and $Fe_{15}$, the diameters of the cluster and the

SWNT are similar. This is due to the strong attraction between the Fe cluster atoms and the bond-unsaturated C atoms at the graphitic cap edges. As illustrated in Fig. 1, the size of the graphitic cap that lifts off the cluster surface is typically smaller than the catalyst diameter (at least for clusters larger than 0.6-0.7 nm), but the cap diameter increases as carbon atoms are incorporated into the structure (Figs 1C-E). The diameter of the cap continues to increase since this leads to an increase in the number of bonds between the cluster Fe atoms and the cap edge atoms. Since the diameter of the SWNT cannot be larger than the cluster diameter it remains constant once it has attained this maximum diameter, and only the length of the SWNT continues to increase as more C atoms are incorporated into the structure.

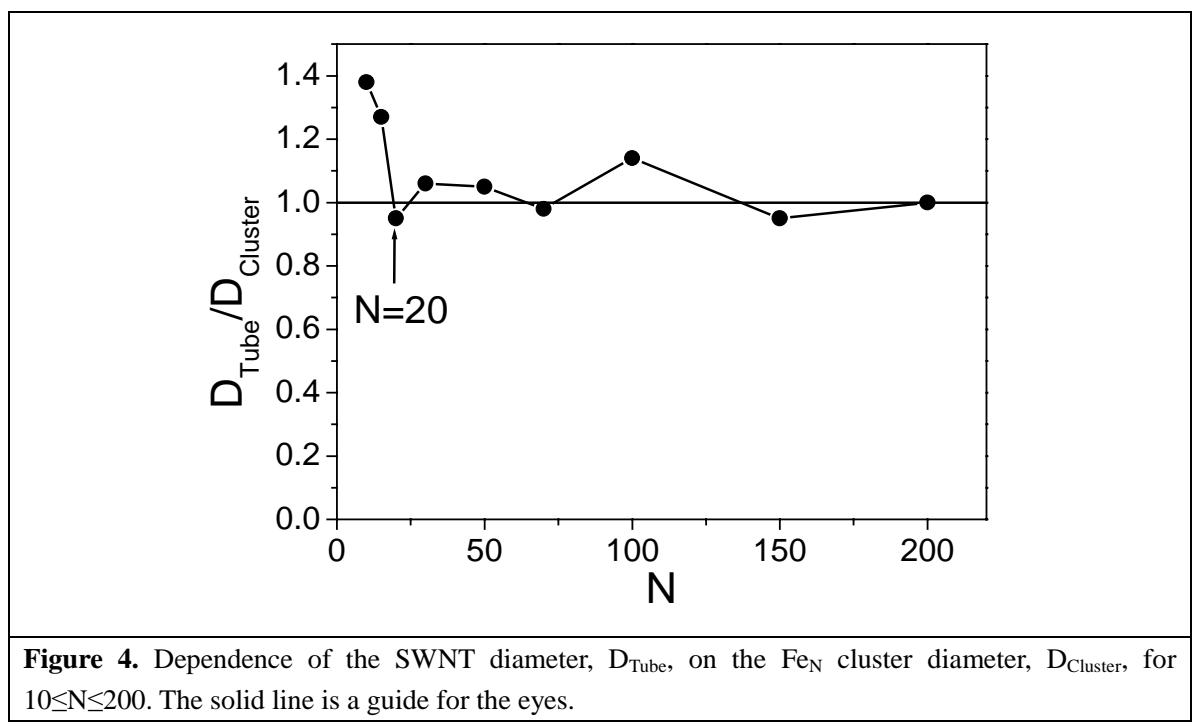

**Figure 4.** Dependence of the SWNT diameter, $D_{Tube}$, on the $Fe_N$ cluster diameter, $D_{Cluster}$, for $10 \leq N \leq 200$. The solid line is a guide for the eyes.

## 4. Conclusion

MD simulations, based on an empirical PES, have been used to study the effect of iron cluster size on the mechanism of catalyzed SWNT growth and on the structure and diameter of the SWNT. The growth mechanism is similar for all cluster sizes studied (the clusters contained between 10 and 200 Fe atoms), and the simulated SWNTs nucleate at temperatures similar to those used in CCVD experiments. Nucleation begins when carbon atoms precipitate to the surface of a highly supersaturated iron-carbide particle. These atoms nucleate carbon strings and polygons that grow into larger graphitic islands, which lift off the surface to form graphitic caps. For large particles, containing at least 20 Fe atoms, the caps grow in diameter until they have the same diameter as the cluster. In this way well-structured SWNTs that have diameters similar to the metal cluster diameters are grown. Also, since the $Fe_N$ cluster is ≈0.6 nm in diameter, the thinnest well-structured SWNT that is grown also has this diameter. Smaller Fe clusters, that have 10 or 15 atoms, also grow SWNTs that are 0.6-0.7 nm in diameter (*i.e.*, the SWNT diameter is larger than the cluster diameter), but these nanotubes have very poor structures that contain many defects and bond-unsaturated carbon atoms. As discussed above, these findings are in agreement with experimental results.

## Acknowledgements

The authors are grateful for the time allocated on the Swedish National Supercomputing facilities and for financial support from the Swedish Research Council, the Swedish Foundation for Strategic Research (CARAMEL consortium) and the Royal Society of Arts and Sciences in Göteborg.

# References

[1] S.Iijima, Nature **354,** 56 (1991).

[2] A. C. Dillon, K. M. Jones, T. A. Bekkedahl, C. H. Klang, D. S. Bethune and M. J. Heben, Nature **386,** 377 (1997).

[3] M. S.Fuhrer, J. Nygard, L. Shih, M. Forero, Y. G. Yoon, M. S. C. Mazzone,
H. J. Choi, J. Ihm, S.G. L ouie, A. Zettl, P. L . McEuen, Science **288**, 494 (2000).

[4] J. Kong, N. R. Franklin, C. Zhou, M. G. Chapline, S. Peng, K. Cho, H. Dai, Science **287,** 622 (2000).

[5] C. Singh, M. S. P. Shaffer, K. K. K. Koziol, I. A. Kinloch, A. H. Windle, Chem. Phy. Lett. **372,** 60 (2003).

[6] Y. Murakami, Y. Miyauchi, S. Chiashi, S. Maruyama, Chem. Phy. Lett. **377**, 49 (2003).

[7] HW. Zhu, CL. Xu, DH. Wu, BQ. Wei, R. Vajtai, PM. Ajayan, Science **296,** 884 (2002).

[8] S. Huang, X. Cai, and J. Liu, J. Am. Chem. Soc. **125,** 5636 (2003).

[9] J. Kong, H. Soh, A. Cassell, H. Dai, Nature **395,** 878 (1998).

[10] K. Bladh, L. K. L. Falk, F. Rohmund, Applied Phys A **70,** 317 (2000).

[11] Y. Saito *et al.,* Jpn. J. Appl. Phys. **33**, 526 (1994); Carbon **33,** 979 (1995).

[12] R. T. K. Bker, M. A. Barber, P. S. Harris, F. S. Feates, and R. J. Waite, J. Catal. **26,** 51 (1972).

[13] J. Gavillet, A. Loiseau, C. Journet, F. Willaime, F. Ducastelle, and J. C. Charlier, Phys. Rev. Lett. **87,** 275504 (2001).

[14] Y. Shibuta, S. Maruyama, Physica B **323,** 187 (2002).

[15] F. Ding, A. Rosén and K. Bolton, submitted.

[16] F. Ding, K. Bolton, and A. Rosén, J. Vac. Sci. Technol. A, accepted.

[17] D.W. Brenner, Phys. Rev. B **42**, 9458 (1990).

[18] V. Rosato, M. Guillope, and B. Legrand, Philos. Mag. A **59**, 321 (1989).

[19] L. J. Lewis, P. Jensen, and J-L. Barrat, Phys. Rev. B **56**, 2248 (1997).

[20] J. Stanek, G. Marest, H. Ja.rezic, and H. Binczycka, Phys. Rev. B **52**, 8414( 1995).

[21] R. A. Johnson, Phys. Rev. **134,** A1329 (1964).

[22] A. V. Evteev, A. T. Kosilov and E. V. Levtchenko, Proceeding of the 22nd Riso international Symposium on Material Science: (2001).

[23] E. Durgun, S. Dag, V. M. K. Bagci, O. Gulseren, T. Yildirim, and S. Ciraci, Phys. Rev. B **67**, 201401 (2003).

[24] G. L. Gutsev, C. W. Bauschlicher Jr., Chem. Phys. **291,** 27 (2003).

[25] H. J. C. Berendsen, J. P. M. Postma, W. F. van Gunsteren, A. DiNola, J. R. Haak, J. Chem. Phys. **81,** 3684 (1984).

[26] S. Y. Chen, H. Y. Miao, J. T. Lue, and M. S. Ouyang, J. Phys. D: Appl. Phys. **37,** 273 (2004).

[27] A. M. Cassell, J. A. Raymakers, J. Kong, and H. Dai, J. Phys. Chem. B **103** 6484 (1999).

[28] S. Maruyama, Y. Miyauchi, T. Edamura, Y. Igarashi, S. Chiashi, Y. Murakami, Chem. Phys. Lett. **375,** 553 (2003).

[29] H. J. Jeong, K. H. An, S. C. Lim, M.-S. Park et. al., Chem. Phy. Lett. **380,** 263 (2003).

[30] N. Wang, Z. K. Tang, G. D. Li and J. S. Chen, Nature **408**, 50 (2000).

[31] D. H. Robertson, D. W. Brenner, and J. W. Mintmire, Phys. Rev. B **45,** 12592 (1992).

[32] D. Ciuparu, Y. Chen, S. Lim, G. L. Haller and L. Pfefferle, J. Phys. Chem. B **108,** 503 (2004).